\documentclass[11pt]{article}

\usepackage[margin=1in]{geometry}

\usepackage{amsthm,amsmath,amsfonts,amssymb}
\usepackage{bm}
\usepackage{graphicx}
\usepackage{rotating}
\usepackage{float}
\usepackage{enumerate,enumitem}
\usepackage{multirow}
\usepackage{verbatim}
\usepackage[font=footnotesize]{subcaption}
\usepackage{tikz}
\usepackage{booktabs}
\usepackage[normalem]{ulem}
\usepackage{algorithm}
\usepackage{algorithmic}
\usepackage{changepage}
\usepackage{threeparttable}
\usepackage{upgreek}
\usepackage[table]{xcolor}
\usepackage[english]{babel}
\usepackage{url}

\usepackage[authoryear]{natbib}
\usepackage[colorlinks=true,citecolor=blue,urlcolor=blue,linkcolor=blue]{hyperref}

\theoremstyle{plain}

\newtheorem{theorem}{Theorem}[section]

\newtheorem{corollary}[theorem]{Corollary}

\theoremstyle{definition}

\title{Enhancing Inference for Small Cohorts via Transfer Learning and Weighted Integration of Multiple Datasets}

\author{
Subharup Guha\\
Department of Biomedical Data Science, Dartmouth College, Hanover, U.S.A.\\
\texttt{subharup.guha@dartmouth.edu}
\and
Mengqi Xu\\
Department of Statistics and Actuarial Science, University of Waterloo, Waterloo, Ontario, Canada\\
\texttt{m332xu@uwaterloo.ca}
\and
Yi Li\\
Department of Biostatistics, University of Michigan, Ann Arbor, Michigan, U.S.A.\\
\texttt{yili@umich.edu}
}

\date{}

\begin{document}

\maketitle

\begin{abstract}
Lung sepsis remains a significant clinical concern in the Northeastern U.S., yet patients from this region are underrepresented in national databases such as the eICU Collaborative Research Database. This underrepresentation limits reliable inference on key clinical markers, FiO$_2$, creatinine, platelets, and lactate, that reflect oxygenation, kidney function, coagulation, and metabolism, and may differ by sex. To address this challenge, we introduce \underline{trans}fer \underline{l}e\underline{a}rning wi\underline{t}h w\underline{e}ights (\textsc{Translate}), a novel adaptive weighting framework for statistical transfer learning across multi-cohort observational data. 
\textsc{Translate} learns cohort-specific weights that account for both cohort prevalence and alignment with a small anchor cohort, enabling data-driven, estimand-agnostic integration of information across dissimilar external cohorts. These weights reflect each cohort's effective contribution to an anchor-aligned pseudopopulation and naturally downweight non-comparable cohorts. The method generalizes existing weighting strategies, offers theoretical guarantees for improved precision, and supports a broad class of estimands, including means, variances, distribution functions, covariances, and pairwise correlations. 
Simulation studies and a real-data application to lung sepsis show that \textsc{Translate} yields more accurate and robust inference for small anchor cohorts by effectively leveraging information from diverse external cohorts.
\end{abstract}



\section{Introduction}\label{S:intro}

Lung sepsis remains a critical health concern, particularly in the Northeastern United States, where environmental exposures, healthcare access, and demographic factors may  result in  disproportionately high mortality rates \citep{seymour2017epidemiology}. Moreover, sex-based differences may contribute to distinct histopathological mechanisms; for instance, men have a higher incidence of sepsis, potentially due to differences in immune response, hormone regulation, inflammatory pathways, and physiological traits  \citep{angele2014gender}. 

 We analyzed the national eICU Collaborative Research Database, focusing on lung sepsis patients from intensive care units across the United States \citep{pollard2018eicu}. The cohort included 408 patients from the Northeast, 3,312 from the Midwest, 1,551 from the South, and 1,695 from the West. Key outcomes included FiO$_2$, creatinine, platelets, and lactate, representing oxygenation, kidney function, coagulation, and metabolism \citep{west2012respiratory}, along with demographic variables such as age, sex, ethnicity, and APACHE Score \citep{zimmerman2006apache}; see Table~S1 in the Supplementary Material. These variables showed regional variability, and analyzing sex-based differences highlights factors important for precision assessment across subgroups.

  When data are limited for a \textit{target} or \textit{anchor cohort}, such as the small Northeast group, drawing reliable inferences is difficult. Incorporating information from larger or related \textit{source} or \textit{external cohorts} can improve estimation accuracy. \textit{Transfer learning} provides a principled framework for this purpose \citep{weiss2016survey}, leveraging external knowledge to refine inference, enhance prediction, and yield more generalizable insights \citep{dahabreh2019extending}. 
 Several statistical approaches have been developed for transfer learning. Likelihood-based methods \citep{chatterjee2016, huang2016efficient} incorporate external data into regression models but rely on strong assumptions about cohort comparability, while adaptive procedures \citep{chen2021combining} use aggregate information to mitigate cross-study heterogeneity. Recent extensions demonstrate the versatility of transfer learning in graphical models, high-dimensional regression, classification, generalized linear models, survival analysis, and federated settings \citep{li2023transfer, li2022transfer, cai2021transfer, tian2023transfer, li2023accommodating}. Weighting-based strategies provide a unifying framework for data integration \citep{borenstein2009introduction}. In particular, importance weighting methods \citep{sugiyama2008direct} correct for covariate shift by reweighting external samples to match the covariate distribution of the anchor cohort, while hierarchical partial pooling approaches \citep{bates2015fitting, mcelreath2018statistical} borrow strength across cohorts by modeling outcome-level heterogeneity through shared random effects and regularized shrinkage \citep{he2013evaluating}.

However, most methods are  tailored to specific univariate outcomes and lack the flexibility to handle  study-specific heterogeneity in cohort design, covariate distributions,  multivariate outcomes and their functionals \citep{adhikari2021data,tax2025multicenter},  increasing the risk of degraded inferences and negative transfer \citep{torrey2010transfer}.
  Established approaches, such as importance weighting and hierarchical partial pooling, also face challenges: importance weighting adjusts for covariate differences but struggles with outcome heterogeneity and unstable weights in high-dimensional or small-sample settings \citep{kimura2024short}, while hierarchical partial pooling assumes shared patterns and may fail to correct covariate shifts \citep{cran2023pooling}.  There is a need for more robust methods that simultaneously address covariate and outcome heterogeneity, particularly for multivariate outcomes and complex population structures \citep{bayer2022accommodating}.

To overcome these limitations, we develop a transfer learning framework that synthesizes evidence across heterogeneous sources using adaptive, domain-specific weights. Unlike fixed- or random-effects meta-analyses \citep{hedges1998fixed}, which impose homogeneity or prespecified hierarchies, our method adapts to covariate and outcome heterogeneity across studies by constructing dynamically weighted \textit{anchor-aligned pseudopopulations} calibrated to the smaller anchor cohort. These subject-specific weights enable unbiased, robust inference on key features, such as means, variances, covariances, and correlations, and their subgroup comparisons within the anchor cohort. 

Within this framework, we construct the \underline{trans}fer \underline{l}e\underline{a}rning wi\underline{t}h w\underline{e}ights (\textsc{Translate}) population, which maximizes the composite effective sample size (ESS) and enables flexible, accurate estimation of complex multivariate functionals without prespecifying a fixed estimand. Theoretical results show that \textsc{Translate} achieves higher precision than analyses based solely on the anchor cohort by assigning cohort probabilities proportional to their ESS, thereby down-weighting external cohorts that differ substantially from the anchor. Unlike general anchor-aligned pseudopopulations, the \textsc{Translate} framework ensures additive cohort contributions to the composite ESS. We further develop new techniques for integrative inference and establish the asymptotic properties of the resulting multivariate estimators.

The remainder of the paper is organized as follows. Section~\ref{S:method} presents the methodology through a two-stage framework and examines its theoretical properties. In Stage~1, we introduce the anchor-aligned pseudopopulation, which learns subject-specific weights from covariate and outcome differences across cohorts; \textsc{Translate}, a specific implementation, is then described. Stage~2 develops weighted estimators for outcome functionals. Section~\ref{S:simulation} evaluates the method on simulated data, and Section~\ref{S:data analysis} applies it to lung sepsis outcomes in the Northeastern U.S., integrating 6,966 ICU patients to improve inference for the small regional cohort and identify sex-based differences in creatinine and platelet function. Section~\ref{S:discussion} concludes the paper. Proofs appear in the Supplementary Material.


\section{\underline{Trans}fer \underline{L}e\underline{a}rning Wi\underline{t}h W\underline{e}ights (\textsc{Translate})}\label{S:method}

 We consider a medical study involving a small anchor cohort (indexed by $s = 0$) with $N_0$ subjects, $p$ covariates, and $L$ outcomes, where $p$ and $L$ are of moderate dimension. The goal is to estimate outcome functionals such as means, medians, variances, cumulative distribution functions (CDFs), covariances, and pairwise correlations, and to compare them across subgroups (e.g., by sex or treatment). Motivated by the limited precision achievable within the small anchor cohort, we propose a transfer learning framework that incorporates auxiliary information from $J$ larger external cohorts, where $J$ is moderate and may be as small as one.  
For $s = 1, \ldots, J$, the $s$th external cohort contains $N_s \gg N_0$ subjects with the same medical condition and records the same subject-level covariates and outcomes. Let $s_i \in \{0, \ldots, J\}$ denote the cohort label, $\mathbf{x}_i \in \mathcal{X} \subset \mathbb{R}^p$ the covariate vector, and $\mathbf{Y}_i \in \mathcal{Y} \subset \mathbb{R}^L$ the observed outcomes for subject $i = 1, \ldots, N$.    

In the motivating lung sepsis application, the dataset comprises $N = 6{,}966$ patients, with $N_0 = 408$ from the Northeastern region serving as the anchor cohort and the remaining patients distributed across $J = 3$ external cohorts. Each patient has $p = 46$ demographic and admission-related covariates and $L = 4$ clinical outcomes: FiO$_2$, creatinine, platelet levels, and lactate.  

If the subject labels are randomly assigned,  the $N$ subject-specific cohort labels, covariates and outcomes are exchangeable and can be regarded as an i.i.d.\ sample from an   \textit{composite population},  $f(S,\mathbf{X},\mathbf{Y})$, where  $f(\cdot)$ generically represents distributions or densities. 
  The prevalence of the $s$th cohort, $\pi_s$, is defined as the  probability of  event $[S = s]$ under the composite population and is estimated  by $\hat{\pi}_s = N_s / N$. For the motivating data, these estimates are $\hat{\pi}_0=0.06$, $\hat{\pi}_1=0.48$, $\hat{\pi}_2=0.22$, and $\hat{\pi}_3=0.24$. 
  %
  %
  In the composite population, denote the covariate density and  conditional response density in cohort $s$ by $f^{(s)}_{\mathbf{X}}(\mathbf{x})$ and $f^{(s)}_{\mathbf{Y} \mid \mathbf{X}}(\mathbf{y}; \mathbf{x})$, respectively. Let $\mathbf{Z} := (\mathbf{X}, \mathbf{Y})$. For  $(s,\bm{z}) \in  \boldsymbol{\Omega} = \{0,\ldots,J\} \times \mathcal{X} \times \mathcal{Y}$,
  \begin{align}
    f^{(s)}_{\mathbf{Z}}(\bm{z}) 
    &= f^{(s)}_{\mathbf{X}}(\mathbf{x}) f^{(s)}_{\mathbf{Y} \mid \mathbf{X}}(\mathbf{y}; \mathbf{x}), \quad\text{and additionally,}
    \label{XY}\\[4pt]
    f(s,\bm{z})  
    &= \pi_s \, f^{(s)}_{\mathbf{Z}}(\bm{z}), 
    \label{Dr}
\end{align}
 the composite population  density of $(S,  \mathbf{Z})$.  
The   density of $\mathbf{Z}$ in the composite population, irrespective of cohort, is then
\(
f_{\mathbf{Z}}(\bm{z})  = \sum_{s=0}^J \pi_s f^{(s)}_{\mathbf{Z}}(\mathbf{z}).
\)

\begin{figure}[htbp]
  \centering
  \includegraphics[width=\textwidth]{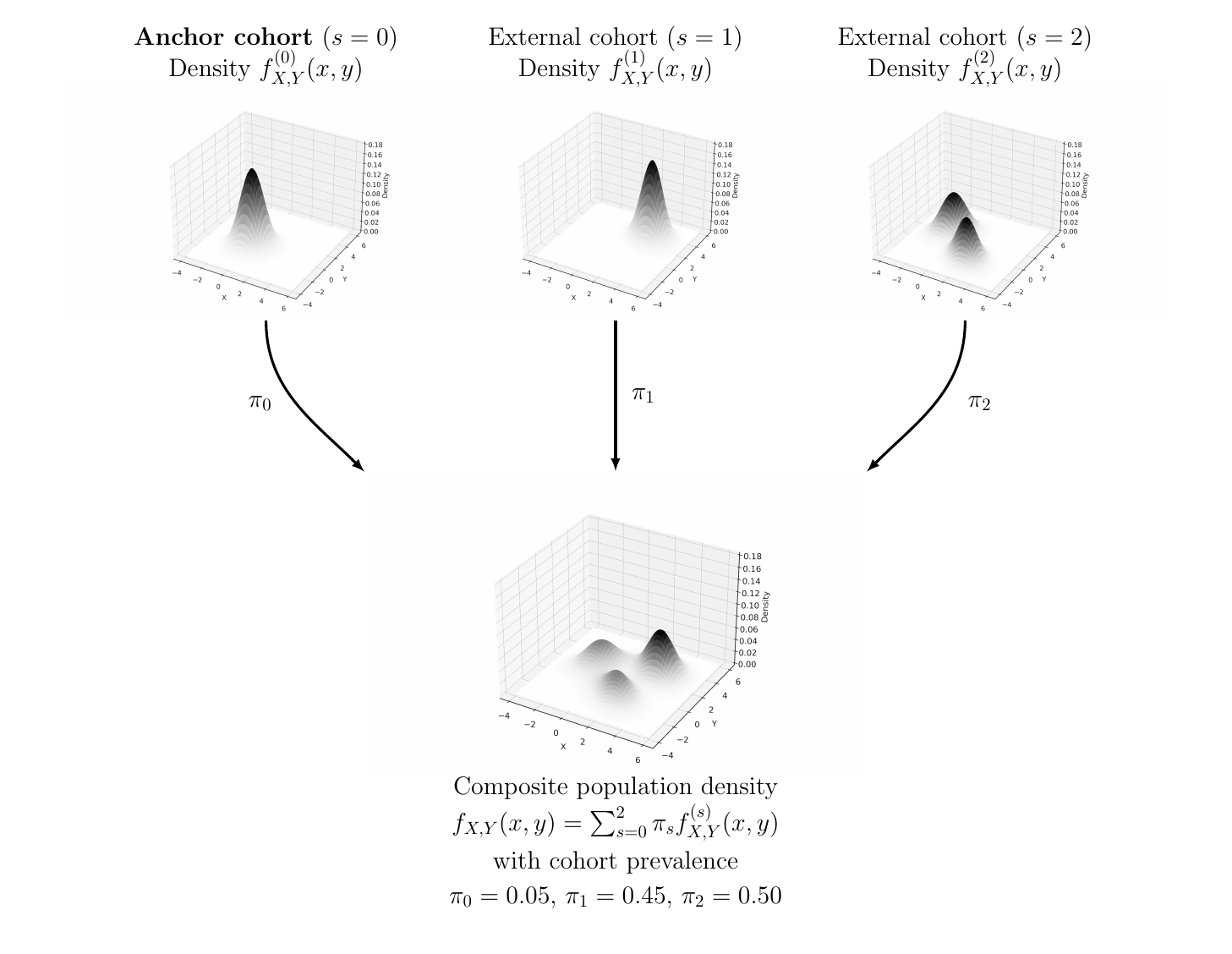}
  \caption{\textbf{Illustration of cohort prevalence and composite population.}
This figure uses a toy example with univariate, continuous covariates and outcomes to illustrate the concept of composite population in a multi-cohort observational study.
The top row shows the joint distributions $f^{(s)}_{X,Y}(x,y)$ for the anchor cohort ($s=0$) and two external cohorts ($s=1,2$), each exhibiting distinct $(x,y)$ characteristics.
The bottom panel displays the resulting composite population, formed by combining the individual cohorts with prevalences $\pi_0 = 0.05$, $\pi_1 = 0.45$, and $\pi_2 = 0.50$.
This mixture, denoted by $f_{X,Y}(x,y) = \sum_{s=0}^2 \pi_s f^{(s)}_{X,Y}(x,y)$, represents the distribution from which the observed $(x_i, y_i)$ values can be regarded as $N$ i.i.d. realizations.
While the figure uses simple univariate continuous variables for visualization, the framework naturally extends to multivariate and mixed-type covariates and outcomes.
}
  \label{fig:1a}
\end{figure}

To aid visualization, Figure~\ref{fig:1a} uses a toy example with univariate, continuous covariates $X$ and outcomes $Y$. The anchor cohort ($s = 0$) and two external cohorts ($s = 1, 2$) exhibit distinct joint distributions $f^{(s)}_{X,Y}(x, y)$, shown in the top row. The bottom panel shows the composite population formed by mixing these cohorts using theoretical prevalences $\pi_0 = 0.05$, $\pi_1 = 0.45$, and $\pi_2 = 0.50$. 
The resulting density,
\(
f_{\mathbf{Z}}(\mathbf{z}) = f_{X,Y}(x,y) = \sum_{s=0}^2 \pi_s f^{(s)}_{X,Y}(x,y),
\)
represents the population from which the covariates and outcomes of the $N$ subjects can be regarded as i.i.d.\ realizations.

We aim to leverage all $N = \sum_{s=0}^J N_s$ subjects to improve inferential precision for the smaller anchor cohort. 
The \textit{naive combination} strategy  pools all subjects into a single super-cohort of size $N$, but the resulting inferences pertain to the composite population (bottom panel of Figure~\ref{fig:1a}) rather than the anchor cohort (upper-left panel), leading to bias since the anchor cohort is the target population. This bias is amplified when the covariate and outcome distributions of the larger external cohorts differ substantially from those of the anchor. The same reasoning extends naturally to multivariate and mixed-type data settings.

\subsection{Stage 1: Evaluating sample weights} \label{S:stage 1}

Suppose we integrate the $(J+1)$ cohorts by assigning normalized subject weights $\tilde{w}_i$, forming a weighted sample $\{(\tilde{w}_i,\mathbf{z}_i): i=1,\ldots,N\}$ that flexibly adjusts for covariate and outcome differences between external and anchor cohorts and yields unbiased, more precise anchor-cohort inference than either anchor-only analysis or naive pooling.

\subsubsection{Anchor-aligned  pseudopopulations}\label{S:general Stage 1} We first realign the covariate-outcome relationships within each external cohort using weights that transform the composite population into a hypothetical \textit{anchor-aligned pseudopopulation} mirroring the  anchor cohort within each external cohort.
 Weighted inferences based on all $N$ subjects are then unbiased for this pseudopopulation and, by design, enable valid information transfer  for the smaller anchor cohort. 
More formally, the anchor-aligned pseudopopulation is constructed to satisfy the following properties: \textit{(a)} $S \perp \mathbf{Z}$, and \textit{(b)} within each cohort, the distribution of $\mathbf{Z}$ matches the observed distribution in the anchor cohort. For a parameter vector of \textit{alignment proportions} $\boldsymbol{\gamma} = (\gamma_0, \ldots, \gamma_J)$ satisfying $\sum_{s=0}^J \gamma_s = 1$, the joint density in this pseudopopulation is
\begin{equation}
f_{\boldsymbol{\gamma}}(s, \bm{z}) = \gamma_s \, f^{(0)}_{\mathbf{Z}}(\bm{z}), \label{pseudo}
\end{equation}
and $\gamma_s$ represents the pseudopopulation probability of the $s$th cohort. The subscript $\boldsymbol{\gamma}$ indicates anchor-aligned distributions in (\ref{pseudo}), whereas superscript $(0)$ refers to the anchor cohort. Unlike $f(\cdot)$ in (\ref{Dr}), these depend on the design parameter $\boldsymbol{\gamma}$  indexing a family of pseudopopulations satisfying conditions \textit{(a)}-\textit{(b)}. Section \ref{S:translate}   describes a data-driven strategy for selecting $\boldsymbol{\gamma}$ to optimize inference for the anchor cohort.

\paragraph{Adaptive alignment of external cohorts}  Assume mutual absolute continuity of the $(J+1)$ cohort-specific densities, so that $f^{(s)}_{\mathbf{Z}}(\mathbf{z}) > 0$ for all $s$, $\mathbf{x} \in \mathcal{X}$, and $\mathbf{y} \in \mathcal{Y}$. The \textit{cohort alignment factor} is defined as
\begin{align}
    \psi_s(\mathbf{z}) &:= \frac{f^{(0)}_{\mathbf{Z}}(\mathbf{z})}{f^{(s)}_{\mathbf{Z}}(\mathbf{z})}, \quad (s, \mathbf{z}) \in \boldsymbol{\Omega}. \label{psi}
\end{align}
Since $f^{(0)}_{\mathbf{Z}}(\mathbf{z}) = \psi_s(\mathbf{z}) \, f^{(s)}_{\mathbf{Z}}(\mathbf{z})$, the alignment factor redistributes the  density of $\mathbf{Z}$ within each external cohort to match that of the anchor cohort. For the anchor itself, $\psi_0(\mathbf{z}) = 1$ for all $\mathbf{z} = (\mathbf{x}, \mathbf{y})$. Although $\psi_s(\mathbf{z})$ integrates to one within each cohort, its pointwise values may deviate substantially from one in regions where the external density diverges from that of the anchor, resulting in high variability over $\mathcal{X} \times \mathcal{Y}$ for poorly aligned cohorts. 

\begin{sidewaysfigure}
  \centering
  \includegraphics[width=0.9\textheight]{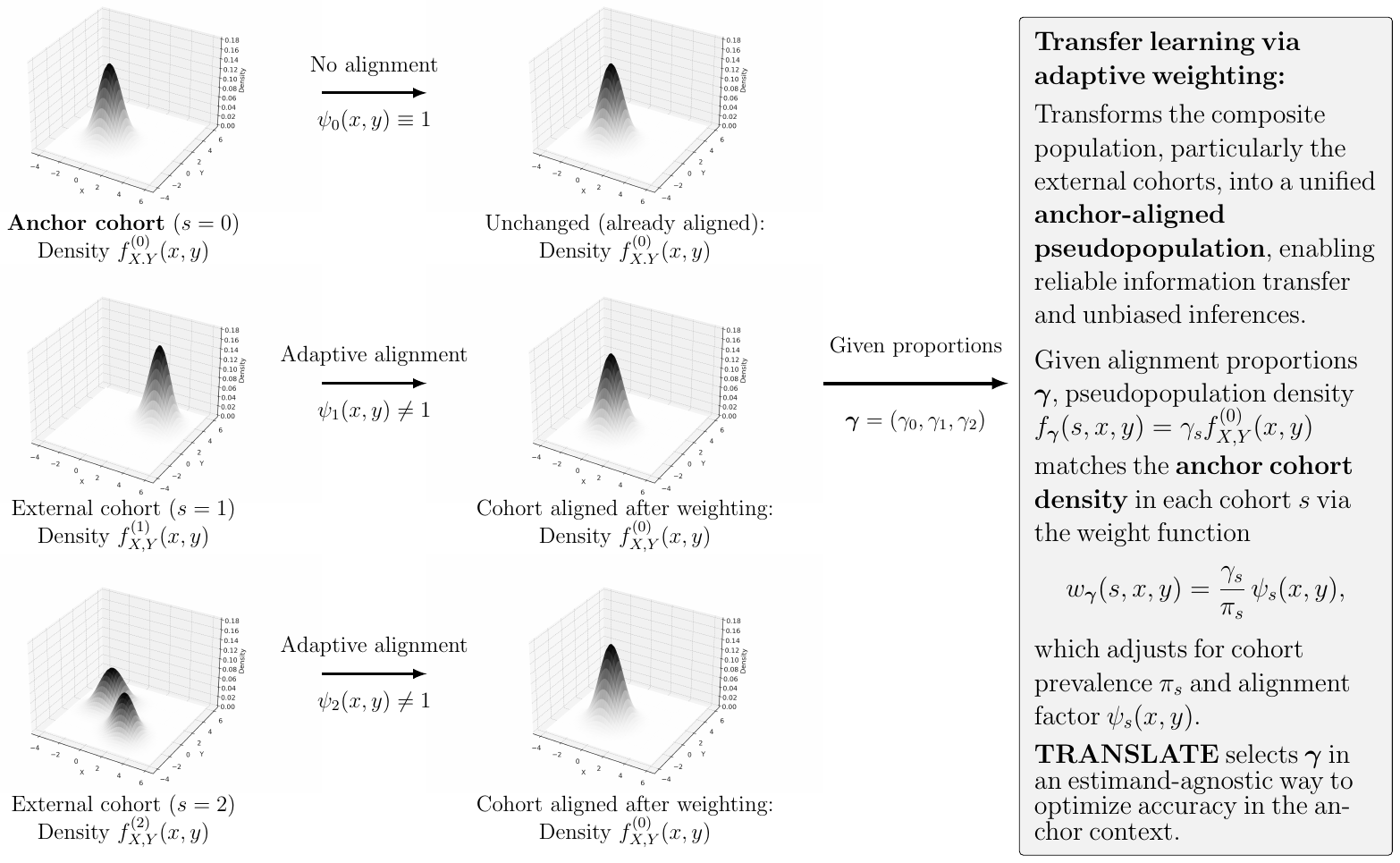}
  \caption{\textbf{Illustration of the \textsc{TRANSLATE} framework using a toy example with $J = 2$ external cohorts.} 
  The horizontal layout demonstrates how adaptive weighting transforms the composite population of Figure~\ref{fig:1a}, especially the external cohorts, into an anchor-aligned pseudopopulation that matches the covariate-outcome distribution of the anchor cohort. This alignment enables unbiased inferences and effective information transfer across dissimilar cohorts. Although the figure uses univariate continuous variables and two external cohorts for visualization, the framework extends to multivariate and mixed-type variables and to any small or moderately large number of external cohorts.}
  \label{fig:1b}
\end{sidewaysfigure}

For the toy example in Figure~\ref{fig:1a}, column 1 of Figure~\ref{fig:1b} shows the joint densities of $\mathbf{Z} = (X, Y)$ for the anchor and external cohorts, highlighting differences in their covariate-outcome distributions. Column 2 shows the alignment factor $\psi_s(x, y)$, depicted by arrows, which reweights each external density $f^{(s)}_{X,Y}(x, y)$ to approximate the anchor density $f^{(0)}_{X,Y}(x, y)$; values far from 1 indicate stronger divergence.

Building on this, we introduce the \textit{alignment weight function}, which transforms  the composite population to the anchor-aligned pseudopopulation for any given set of alignment proportions $\boldsymbol{\gamma}$. Using equations~(\ref{Dr})-(\ref{psi}) together with the cohort prevalences $\boldsymbol{\pi}$, the alignment weight for each $(s, \mathbf{z}) \in \boldsymbol{\Omega}$ is defined as
\begin{align}
    w_{\boldsymbol{\gamma}}(s, \mathbf{z}) := \frac{f_{\boldsymbol{\gamma}}(s, \mathbf{z})}{f(s, \mathbf{z})}
    = \frac{{\gamma_s \, f^{(0)}_{\mathbf{Z}}(\mathbf{z})}}{{\pi_s \, f^{(s)}_{\mathbf{Z}}(\mathbf{z})}} 
    = \frac{\gamma_s}{\pi_s} \, \psi_s(\mathbf{z}). \label{w}
\end{align}

Since $f_{\boldsymbol{\gamma}}(s, \mathbf{z}) = w_{\boldsymbol{\gamma}}(s, \mathbf{z}) \, f(s, \mathbf{z})$, the alignment weight function redistributes the composite population density $f(s, \mathbf{z})$ across cohorts and covariate-outcome profiles,  adaptively transforming the composite population into the anchor-aligned pseudopopulation. The last column of Figure~\ref{fig:1b} illustrates this transformation.

\paragraph{Relating cohort alignment factors to cohort-label regression}  
In practice, computing the alignment weights directly from (\ref{psi}) can be computationally intensive and error-prone, as it requires estimating joint multivariate densities of covariates and outcomes within each cohort.
 An alternative, more practical approach leverages Bayes' theorem. Letting $\theta_s(\bm{z}) = f\bigl(S = s \mid \mathbf{Z} = \bm{z}\bigr)$, we obtain an equivalent expression for the cohort alignment factor:
\begin{equation}
    \psi_s(\mathbf{x}, \mathbf{y}) = \frac{\pi_s}{\pi_0} \cdot \frac{\theta_0(\bm{z})}{\theta_s(\bm{z})}, \quad (s, \bm{z}) \in \boldsymbol{\Omega}. \label{psi2}
\end{equation}
Designating the anchor cohort ($s = 0$) as the reference category, we define the log ratio $\eta_s(\bm{z}) = \log\bigl(\theta_s(\bm{z}) / \theta_0(\bm{z})\bigr)$, implying that $\eta_0(\bm{z}) = 0$.
Inferentially, this approach simplifies (\ref{psi}) by substituting multivariate density estimation with categorical regression of cohort labels on covariates and outcomes to estimate cohort probabilities. It can be implemented using either parametric or nonparametric methods, including multinomial logistic regression, random forests \citep{breiman}, BART \citep{chipman2010}, and boosting \citep{Hothorn_Buhlmann_2006}. Nonparametric approaches, in particular, offer flexibility in modeling $\theta_s(\bm{z})$, especially in high-dimensional settings, and often yield sparse, robust approximations.

\subsubsection{\textsc{TRANSLATE}  pseudopopulation}\label{S:translate} 

Each choice of the design parameter $\boldsymbol{\gamma}$ defines a distinct anchor-aligned pseudopopulation in which  the  covariate-response relationships in the external cohorts resemble those observed in the anchor cohort,  yielding valid inferences for that cohort. 
One possible data-driven strategy is to select $\boldsymbol{\gamma}$  to optimize the estimation accuracy of a prespecified feature of the multivariate outcomes. However, this approach may be inadequate when dealing with complex multivariate structures (whether continuous, categorical, or mixed-type) and a broad range of functionals, such as means, pairwise correlations,  subgroup-specific summaries, or any post hoc functionals of interest.

 Motivated by these considerations, we instead focus on a widely used criterion for evaluating the reliability of weighting methods:  \textit{effective sample size} (ESS). Let $\mathbb{E}[\cdot]$ and $\mathbb{V}\{\cdot\}$ denote, respectively, theoretical expectation and variance under the composite population defined in~(\ref{Dr}). For a given anchor-aligned pseudopopulation indexed by $\boldsymbol{\gamma}$, the \textit{(composite) ESS} \citep{mccaffrey2013tutorial}   is defined using the  alignment weight function~(\ref{w}) as
\(
    \mathcal{Q}_N(\boldsymbol{\gamma}) := \frac{N}{1 + \mathbb{V}\left\{w_{\boldsymbol{\gamma}}(S, \mathbf{Z})\right\}} = \frac{N}{\mathbb{E}\left[w^2_{\boldsymbol{\gamma}}(S, \mathbf{Z})\right]},
\)
where the second equality follows from the normalization $\mathbb{E}[w_{\boldsymbol{\gamma}}(S, \mathbf{Z})] = 1$. By construction, $\mathcal{Q}_N(\boldsymbol{\gamma}) < N$, and a higher ESS indicates lower variability in the alignment weights, which in turn implies greater inferential efficiency. Specifically, a composite ESS greater than $N_0$ indicates that transfer learning of the external cohorts via weighting achieves greater precision compared to using only the $N_0$ anchor cohort subjects.

  We define the \textit{$s$th cohort ESS}  analogously, but within each individual cohort, using the cohort factor in (\ref{psi}). Specifically, the $s$th (theoretical) cohort ESS is given by
\(
\mathcal{Q}_{N_s}^{(s)} := \frac{N_s}{1 + \mathbb{V}\left\{\psi_s(\mathbf{Z}) \mid S = s\right\}} = \frac{N_s}{\mathbb{E}\left[\psi_s^2(\mathbf{Z}) \mid S = s\right]}.
\)
For the anchor cohort ($s = 0$), we have $\psi_0(\mathbf{Z}) \equiv 1$, so $\mathcal{Q}_{N_0}^{(0)} = N_0$. For each external cohort, $\mathcal{Q}_{N_s}^{(s)} \le N_s$, with equality when $\psi_s(\mathbf{Z})$ has no variability. Thus, a cohort that is highly similar to the anchor will exhibit low variability in $\psi_s(\mathbf{Z})$ and yield a cohort ESS close to $N_s$. In contrast, a cohort that is markedly different from the anchor will show high variability in $\psi_s(\mathbf{Z})$, resulting in a much smaller cohort ESS.


The \textbf{\textsc{Translate}} (TRANSfer LeArning wiTh wEights) pseudopopulation is the anchor-aligned construct that maximizes the composite ESS, with probability vector $\breve{\boldsymbol{\gamma}}$  identifying this optimum: \[
\breve{\boldsymbol{\gamma}} = \arg\max_{\boldsymbol{\gamma} \in \mathcal{S}^{J+1}}\mathcal{Q}_N(\boldsymbol{\gamma}) = \arg\min_{\boldsymbol{\gamma} \in \mathcal{S}^{J+1}} \,\mathbb{E}\left[w^2_{\boldsymbol{\gamma}}(S, \mathbf{Z})\right],
\] where $\mathcal{S}^{J+1}$ denotes the $(J+1)$-dimensional unit simplex in $\mathbb{R}^{J+1}$. 
     In addition to aligning the composite population with the anchor cohort and achieving a high composite ESS, an effective weighting scheme must also prevent negative transfer by down-weighting external cohorts that differ substantially from the anchor. The following Theorem \ref{Thm1} shows that the \textsc{Translate} pseudopopulation satisfies these properties and provides intuitive expressions for $\breve{\boldsymbol{\gamma}}$ and the corresponding maximum composite ESS. 

\begin{theorem}\label{Thm1}
Consider $N$ subjects distributed across $(J+1)$ cohorts, where each subject's cohort label $s_i$, covariate vector $\mathbf{x}_i$, and outcome vector $\mathbf{y}_i$ are drawn i.i.d.\ from a population with joint density $f(s, \mathbf{z})$, as defined in~\eqref{Dr}, with $(s, \mathbf{z}) \in \boldsymbol{\Omega} = \{0, \ldots, J\} \times \mathcal{X} \times \mathcal{Y}$. Let the design parameter $\boldsymbol{\gamma} = (\gamma_0, \ldots, \gamma_J)$ define an anchor-aligned pseudopopulation as in~\eqref{pseudo}, with corresponding alignment weight function~\eqref{w}, composite effective sample size (ESS) $\mathcal{Q}_N(\boldsymbol{\gamma})$, and cohort-specific ESS values $\mathcal{Q}_{N_0}^{(0)}, \ldots, \mathcal{Q}_{N_J}^{(J)}$.
\begin{enumerate}
    \item\label{part0} The composite ESS of a general anchor-aligned pseudopopulation \eqref{pseudo} is given by
    \(
\mathcal{Q}_N(\boldsymbol{\gamma}) = \left( \sum_{s=0}^J 
    \frac{\gamma_s^2}{\mathcal{Q}_{N_s}^{(s)}} \left( \frac{\hat{\pi}_s}{\pi_s} \right) \right)^{-1}.
    \)
    \item\label{part1} For the  \textsc{Translate} pseudopopulation with alignment proportions $\breve{\boldsymbol{\gamma}}$: \[
\breve{\gamma}_s 
\propto \mathcal{Q}_{N_s}^{(s)} \left( \frac{\pi_s}{\hat{\pi}_s} \right), \quad s = 0, \ldots, J.\]
For large $N$,  $
\breve{\gamma}_s \asymp \mathcal{Q}_{N_s}^{(s)}$.
        \item\label{part2} The  \textsc{Translate}  composite ESS is
        \(
\mathcal{Q}_N(\breve{\boldsymbol{\gamma}}) =  \sum_{s=0}^J \mathcal{Q}_{N_s}^{(s)} \left( \frac{\pi_s}{\hat{\pi}_s} \right).
        \)
        Moreover, as $N \to \infty$, 
        \(
\frac{\mathcal{Q}_N(\breve{\boldsymbol{\gamma}})}{\sum_{s=0}^J \mathcal{Q}_{N_s}^{(s)}} \xrightarrow{\text{a.s.}} 1,\) and $\mathcal{Q}_N(\breve{\boldsymbol{\gamma}}) > N_0$ eventually.
    \end{enumerate}
\end{theorem}

The result offers several important insights. \textit{First}, for the \textsc{Translate} pseudopopulation, Part~\ref{part1} of Theorem~\ref{Thm1} shows that the optimal alignment proportions are asymptotically proportional to the cohort-specific ESS values. 
\textit{Second}, external cohorts more similar to the anchor cohort yield higher ESS values, whereas dissimilar ones yield lower ESS. Consequently, \textsc{Translate} naturally downweights less relevant cohorts in the resulting pseudopopulation. 
\textit{Third}, Part~\ref{part2} shows that, unlike general anchor-aligned pseudopopulations of Part~\ref{part0}, the composite ESS under \textsc{Translate} is asymptotically additive across cohorts. As $N$ grows, it surpasses the anchor cohort size $N_0$, ensuring greater precision than analyses based solely on anchor data and safeguarding against negative transfer. 
\textit{Finally}, when all external cohorts are identical to the anchor cohort in covariate and outcome distributions, the alignment factors equal 1, each cohort's ESS equals its sample size, and the alignment proportions reduce to $\breve{\gamma}_s = \pi_s$. In this case, \textsc{Translate} coincides with the naive pooled analysis.

\subsubsection{Stage 1 inference}\label{S: stage 1 inference}

   Using~(\ref{psi2}), the $s$th theoretical cohort ESS can be written as
 \begin{align}
\mathcal{Q}_{N_s}^{(s)} 
&= \frac{N_s}{\mathbb{E}\!\left[\psi_s^2(\mathbf{Z})\mid S=s\right]}
 = \frac{N_s\pi_0^2/\pi_s^2}{\mathbb{E}\!\left[\theta_0^2(\mathbf{Z})/\theta_s^2(\mathbf{Z})\mid S=s\right]}
 = N\pi_0^2 \bigl(\mathbb{E}[g_s(\mathbf{Z},S)]\bigr)^{-1}, \label{mgf}\\[2pt]
&g_s(\mathbf{Z},S) 
:= \exp\!\bigl(-2\eta_S(\mathbf{Z})\bigr)\,\mathcal{I}(S=s), \notag
\end{align}
Here, $\mathcal{I}(\cdot)$ is the indicator function; $s$ is fixed, while $S$ denotes the random cohort label under the composite population.
For the anchor cohort ($s=0$), $\mathbb{E}[g_0(\mathbf{Z},S)] = \pi_0$, giving $\mathcal{Q}_{N_0}^{(0)} = N_0$.  

We then construct subject-specific weights $\tilde{w}_i$ (normalized to have mean one) so that the weighted sample ${(\tilde{w}_i,\mathbf{x}_i,\mathbf{y}_i) : i = 1,\ldots,N}$ adaptively adjusts for cohort discrepancies and enables valid inference for the anchor-aligned pseudopopulation, including the special case of \textsc{Translate}.
The procedure for estimating these weights is summarized in Algorithm 1. The resulting weighted sample can be viewed as an extended dataset in which  each subject $i$ is effectively replicated $\tilde{w}_i$ times, preserving the overall sample size ($\sum_{i=1}^N \tilde{w}_i = N$).
 
 \begin{algorithm}[htbp]
\caption{Estimation of Anchor-Aligned Weights for a Pseudopopulation}
\label{alg:translate}
\begin{algorithmic}[1]
\small
\STATE \textbf{Input:} Data $\{(s_i,\mathbf{x}_i,\mathbf{y}_i)\}_{i=1}^N$, cohort prevalences $\hat{\boldsymbol{\pi}}$
\STATE \textbf{Step 1: Estimate cohort probabilities.} 
Regress cohort labels on covariates and outcomes to obtain 
$\hat{\eta}_{s_i}(\mathbf{z}_i) = \log[\hat{\theta}_{s_i}(\mathbf{z}_i)/\hat{\theta}_{0}(\mathbf{z}_i)]$, 
using multinomial logistic regression, random forests, BART, or boosting.

\STATE \textbf{Step 2: Estimate cohort alignment factors.}
Compute $\hat{\psi}_{s_i}(\mathbf{z}_i) = (\hat{\pi}_{s_i}/\hat{\pi}_0)\exp[-\hat{\eta}_{s_i}(\mathbf{z}_i)]$.

\STATE \textbf{Step 3: Determine alignment proportions.}
\begin{itemize}[leftmargin=3em]
  \item[\textit{Case 1:}] If $\boldsymbol{\gamma}$ is prespecified, set $\hat{\boldsymbol{\gamma}}=\boldsymbol{\gamma}$.
  \item[\textit{Case 2:}] (\textsc{Translate}) 
Estimate each cohort's sample ESS:
\[
\hat{\mathcal{Q}}_{N_s}^{(s)} 
= \frac{N\hat{\pi}_0^2}{\bar{g}_s}, \quad
\bar{g}_s = \frac{1}{N}\sum_{i=1}^N 
    \exp\!\big[-2\hat{\eta}_{s_i}(\mathbf{z}_i)\big]\mathcal{I}(s_i=s).
\]
Then set 
$\hat{\gamma}_s \propto \hat{\mathcal{Q}}_{N_s}^{(s)}$.
\end{itemize}

\STATE \textbf{Step 4: Compute sample weights.}  
$\hat{w}_{\hat{\boldsymbol{\gamma}}}(s_i,\mathbf{z}_i) = (\hat{\gamma}_{s_i}/\hat{\pi}_{s_i})\,\hat{\psi}_{s_i}(\mathbf{z}_i)$.

\STATE \textbf{Step 5: Normalize weights.}  
$\tilde{w}_i = N\hat{w}_{\hat{\boldsymbol{\gamma}}}(s_i,\mathbf{z}_i) / \sum_{r=1}^N \hat{w}_{\hat{\boldsymbol{\gamma}}}(s_r,\mathbf{z}_r)$.

\STATE \textbf{Step 6: Estimate composite ESS.}  
$\hat{\mathcal{Q}}_N(\hat{\boldsymbol{\gamma}}) = N^2 / \sum_{i=1}^N \tilde{w}_i^2$.

\STATE \textbf{Output:} Normalized weights $\{\tilde{w}_i\}$ and estimated composite ESS.
\end{algorithmic}
\end{algorithm}

\subsection{Stage 2: Estimating outcomes features for the anchor cohort}\label{S:stage 2}

  Using the weighted sample $\{(\tilde{w}_i, \mathbf{x}_i, \mathbf{y}_i): i = 1, \ldots, N\}$, we derive unbiased, efficient inferences for the anchor cohort from functionals of $\mathbf{Z} = (\mathbf{X}, \mathbf{Y})$.  
For real-valued functions $\Phi_1, \ldots, \Phi_M$ on $\mathcal{X} \times \mathcal{Y}$, the targets are $\mathbb{E}[\Phi_m(\mathbf{Z}) \mid S = 0]$, representing, for example, means, variances, or other moments of $\mathbf{Y}$ or $\mathbf{Z}$.  
Let $\boldsymbol{\Phi}(\mathbf{Z}) = (\Phi_1(\mathbf{Z}), \ldots, \Phi_M(\mathbf{Z}))' \in \mathbb{R}^M$, and define 
\[
\boldsymbol{\lambda} := \mathbb{E}[\boldsymbol{\Phi}(\mathbf{Z}) \mid S = 0].
\]
For any $h: \mathbb{R}^M \to \mathbb{R}$, we also estimate $h(\boldsymbol{\lambda})$ as a derived feature of the anchor cohort.  
For example, if $Y_1$ and $Y_2$ are quantitative, set $\Phi_1(\mathbf{Z}) = Y_1$, $\Phi_2(\mathbf{Z}) = Y_1^2$, and $\Phi_3(\mathbf{Z}) = Y_1 Y_2$. Then  
$h(t_1, t_2) = \sqrt{t_2 - t_1^2}$ gives $h(\boldsymbol{\lambda}) = \sigma_1^{(0)}$, the standard deviation of $Y_1$ in the anchor cohort,  
and $h(t_1, t_2, t_3) = \sqrt{t_3 - t_1 t_2}$ yields the covariance between $Y_1$ and $Y_2$.

 As another example, let $y_{11}, \ldots, y_{1M}$ form a grid over the support of $Y_1$ and define $\Phi_m(\mathbf{Z}) = \mathcal{I}(Y_1 \le y_{1m})$. Then $h(t_1, \ldots, t_M) = t_m$ yields the anchor cohort CDF of $Y_1$ at $y_{1m}$, and choosing $m^* = \arg\min_m |t_m - 0.5|$ approximates the median.  
Similarly, to compare conditional means of $Y_1$ across $B$ anchor subgroups (e.g., by sex or treatment), define $\Phi_1(\mathbf{Z}) = Y_1 \mathcal{I}(X_1 = b)$ and $\Phi_2(\mathbf{Z}) = \mathcal{I}(X_1 = b)$ for each $b = 1,\ldots,B$. Then $h(t_1, t_2) = t_1 / t_2$ gives $h(\boldsymbol{\lambda}) = \mathbb{E}[Y_1 \mid S=0, X_1=b]$, the subgroup mean of $Y_1$ in the anchor cohort.

Using the empirically normalized   weights $\tilde{w}_1,\ldots, \tilde{w}_N$
 computed in Section \ref{S: stage 1 inference}, 
we estimate  $\boldsymbol{\lambda}$,  a vector of length $M$, as: 
\begin{equation}
\hat{\boldsymbol{\lambda}} = \frac{1}{N}\sum_{i=1}^N \tilde{w}_i\,\mathbf{\Phi}(\mathbf{Z}_i). \label{l.hat}
\end{equation}
Theorem~\ref{Thm:estimation} and Corollary~\ref{corollary 2} establish the asymptotic normality of $\hat{\boldsymbol{\lambda}}$, validating its use for inference on the anchor cohort characteristics represented by $\boldsymbol{\lambda}$.

\begin{theorem}\label{Thm:estimation}
Consider an anchor-aligned pseudopopulation as defined in~\eqref{pseudo}, with alignment weight function $w_{\boldsymbol{\gamma}}(S, \mathbf{Z})$ given in~\eqref{w}. For $m = 1, \ldots, M$, let $\Phi_m$ be a real-valued function defined on $\mathcal{X} \times \mathcal{Y}$, and define the vector-valued function $\boldsymbol{\Phi}(\mathbf{Z}) = \bigl(\Phi_1(\mathbf{Z}), \ldots, \Phi_M(\mathbf{Z})\bigr)'$. Assume that $\mathbb{E}\bigl[w_{\boldsymbol{\gamma}}^2(S, \mathbf{Z})\bigr] < \infty$ and $\mathbb{E}\bigl[w_{\boldsymbol{\gamma}}^2(S, \mathbf{Z}) \, \Phi_m^2(\mathbf{Z})\bigr] < \infty$ for each $m$, where, as before, $\mathbb{E}[\cdot]$ denotes expectation under the composite population in~\eqref{Dr}. Let $\boldsymbol{\lambda} = \mathbb{E}\bigl[\boldsymbol{\Phi}(\mathbf{Z}) \mid S = 0\bigr]$ denote the anchor cohort feature of interest.
The baseline asymptotic variance matrix is defined as
\(
\boldsymbol{D}_{\boldsymbol{\gamma}, \boldsymbol{\pi}, \boldsymbol{\theta}} = \mathbb{E}\left[w_{\boldsymbol{\gamma}}^2(S, \mathbf{Z}) \left\{\boldsymbol{\Phi}(\mathbf{Z}) - \boldsymbol{\lambda} \right\} \left\{\boldsymbol{\Phi}(\mathbf{Z}) - \boldsymbol{\lambda} \right\}' \right],
\)
and it depends on the design parameter $\boldsymbol{\gamma}$, cohort prevalence $\boldsymbol{\pi}$, and the alignment weights through the parameters $\boldsymbol{\theta}=(\theta_0,\ldots,\theta_J)$  as defined in (\ref{psi2}).
Then, the estimator $\hat{\boldsymbol{\lambda}}$ satisfies
\(
\sqrt{N}\bigl(\hat{\boldsymbol{\lambda}} - \boldsymbol{\lambda}\bigr) \xrightarrow{d} \mathcal{N}_M\bigl(\mathbf{0}, \boldsymbol{\Sigma}\bigr),
\)
where the asymptotic variance matrix $\boldsymbol{\Sigma}$ is given by the following cases:
\begin{enumerate}
    \item If the parameters $\boldsymbol{\theta}$ are known, then
    \(
    \boldsymbol{\Sigma} = \boldsymbol{D}_{\boldsymbol{\gamma}, \boldsymbol{\pi}, \boldsymbol{\theta}} + \mathbf{D}_{\boldsymbol{\theta}}(\boldsymbol{\gamma}, \boldsymbol{\pi}),
    \)
    where the adjustment term $\mathbf{D}_{\boldsymbol{\theta}}(\boldsymbol{\gamma}, \boldsymbol{\pi})$, detailed in the Supplementary Material, accounts for the estimation uncertainty in $\boldsymbol{\gamma}$ and $\boldsymbol{\pi}$.

    \item If the parameters $\boldsymbol{\theta}$ are unknown and estimated using multinomial logistic regression, then
    \(
    \boldsymbol{\Sigma} = \boldsymbol{D}_{\boldsymbol{\gamma}, \boldsymbol{\pi}, \boldsymbol{\theta}} + \mathbf{D}_{\boldsymbol{\theta}}(\boldsymbol{\gamma}, \boldsymbol{\pi}) + \mathbf{D}(\boldsymbol{\theta}),
    \)
    where the additional term $\mathbf{D}(\boldsymbol{\theta})$ accounts for the uncertainty in estimating $\boldsymbol{\theta}$. Both adjustment matrices are defined in the Supplementary Material and further discussed in the remarks following the theorem.
\end{enumerate}
\end{theorem}

\begin{corollary}\label{corollary 2}  
 Let $h$ be a real-valued differentiable function defined on the domain $\mathbb{R}^M$.
Define $\boldsymbol{\lambda} \equiv \mathbb{E}\bigl[\boldsymbol{\Phi}(\mathbf{Z}) \mid S = 0\bigr]$, and suppose the gradient vector $\nabla h(\boldsymbol{\lambda}) = \partial h(\boldsymbol{\lambda}) / \partial \boldsymbol{\lambda}$ is non-zero at $\boldsymbol{\lambda}$.
Let $\boldsymbol{\Sigma}$ denote the asymptotic variance matrix defined in Theorem~\ref{Thm:estimation}, and define
\(
\tau^2 = \nabla' h(\boldsymbol{\lambda}) \, \boldsymbol{\Sigma} \, \nabla h(\boldsymbol{\lambda}).
\)
Then, the plug-in estimator $h(\hat{\boldsymbol{\lambda}})$ satisfies the following large-sample distribution:
\(
\sqrt{N} \bigl( h(\hat{\boldsymbol{\lambda}}) - h(\boldsymbol{\lambda}) \bigr) \xrightarrow{d} N(0, \tau^2).
\)
\end{corollary}

 Corollary \ref{corollary 2} follows directly from Theorem \ref{Thm:estimation} via the delta method. Together, they highlight key properties of the estimator (\ref{l.hat}): (i) the results apply to any anchor-aligned pseudopopulation (\ref{pseudo}), including \textsc{Translate}; (ii) $\boldsymbol{\Sigma}$ denotes the overall asymptotic variance of $\hat{\boldsymbol{\lambda}}$, accounting for uncertainty in $\boldsymbol{\gamma}$, $\boldsymbol{\pi}$, and $\boldsymbol{\theta}$; (iii) $\mathbf{D}_{\boldsymbol{\theta}}(\boldsymbol{\gamma},\boldsymbol{\pi})$ adjusts for estimation of $\boldsymbol{\gamma}$ and $\boldsymbol{\pi}$ when $\boldsymbol{\theta}$ is known, whereas $\mathbf{D}(\boldsymbol{\theta})$ captures additional effects from estimating $\boldsymbol{\theta}$ and vanishes if $\boldsymbol{\theta}$ is fixed; (iv) these adjustment matrices need not be positive or negative definite, so in some cases estimation of $(\boldsymbol{\gamma},\boldsymbol{\pi},\boldsymbol{\theta})$ may unexpectedly reduce the asymptotic variance of $h(\hat{\boldsymbol{\lambda}})$; that is, \(
\nabla'h(\boldsymbol{\lambda})\,\boldsymbol{\Sigma}\,\nabla h(\boldsymbol{\lambda}) 
< 
\nabla'h(\boldsymbol{\lambda})\,\boldsymbol{D}_{\boldsymbol{\gamma},\boldsymbol{\pi},\boldsymbol{\theta}}\,\nabla h(\boldsymbol{\lambda}),
\)
although such reduction is not guaranteed for all $h$;   and (v) for large $N$, inference on anchor-cohort characteristics can rely on asymptotic normality with well-behaved standard errors, though bootstrap estimation is used in practice due to the complexity of the variance expressions.


\section{Simulation Study}\label{S:simulation}

We conducted simulations to assess how the proposed weighting methods improve inference for a small anchor cohort using information from a large external cohort. Each of the $R = 250$ datasets included one anchor and one external cohort ($J = 1$), with the anchor comprising $\pi_0 = 0.05$ of $N = 5{,}000$ or $10{,}000$ subjects. Each dataset contained $p = 4$ covariates and a bivariate outcome $\mathbf{Y} = (Y_1, Y_2)'$. Two heterogeneity settings were examined: (i) dissimilar $Y$ only and (ii) dissimilar $X$ and $Y$, producing four total scenarios.

Stage~1 applied the \textsc{Translate} framework  to estimate subject-specific weights via quadratic discriminant analysis (\texttt{qda} in \texttt{MASS}). Stage~2 used these weights to estimate $\mathbb{E}[Y_1 \mid S = 0]$, $\mathbb{S}(Y_2 \mid S = 0)$, and $\mathbb{C}(Y_1, Y_2 \mid S = 0)$ using the weighted estimator~(\ref{l.hat}). We compared three transfer learning strategies: (i) naive pooled (``Naive''), (ii) Importance Weighting (``IW''), (iii) Hierarchical Partial Pooling (``HPP''), and (iv) \textsc{Translate}, against known theoretical anchor values and corresponding effective sample sizes. IW and HPP were implemented via \texttt{survey} and \texttt{lme4}, respectively. For each replication ($r = 1, \ldots, 250$), cohort indicators were generated as $s_i^{[r]} \sim \mathrm{Bernoulli}(0.9)$. Covariates were drawn as 
\[
x_{i1} \sim \mathrm{Bernoulli}(0.2), \quad 
x_{i2} \sim \mathrm{Uniform}(0,0.1), \quad
x_{i3} \sim N(0,0.1), \quad 
x_{i4} \mid s_i \sim N(\phi_x s_i, 0.1),
\]
where $\phi_x = 0$ or $1$ in the ``dissimilar $Y$'' and ``dissimilar $X,Y$'' scenarios, respectively. Conditional on covariates and cohort, outcomes followed  
\[
y_{il} \mid \mathbf{x}_i, s_i \sim N\!\left(\sum_{j=1}^4 x_{ij} + \phi_y s_i, \sigma_{s_i}^2\right), 
\quad l = 1, 2,
\]
with $\phi_y = 1.5$, $\sigma_0 = 0.5$, $\sigma_1 = 0.6$. The conditional covariance was specified as
\[
\mathbb{C}(y_{i1}, y_{i2} \mid \mathbf{x}_i, s_i) = (-1)^{1 + s_i} \, 0.5 \, \sigma_{s_i}^2,
\]
yielding negative within-anchor and positive within-external correlations.

\begin{figure}[t]
\centering
\includegraphics[scale=0.6]{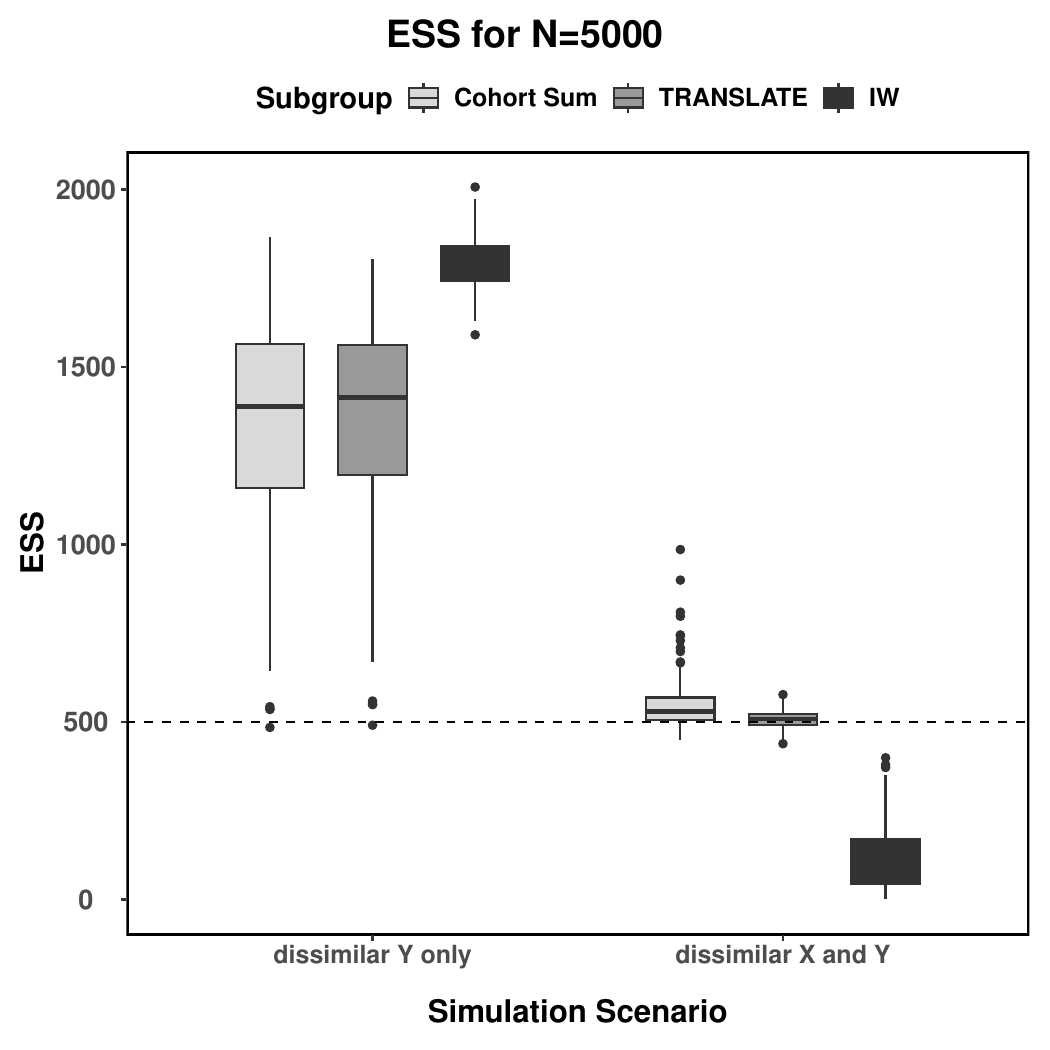}
\caption{Side-by-side boxplots of the composite ESS across 250 simulated datasets for $N = 5{,}000$ under two simulation scenarios, comparing the \textsc{Translate} and \textit{Importance Weighting (IW)} methods and the cohort-level ESS sums. The horizontal line demarcates the expected anchor cohort size, $N\pi_0$. The simulation scenario labeled \textit{dissimilar $Y$ only} corresponds to dissimilar outcomes with similar covariates, and the scenario labeled \textit{dissimilar $X$ and $Y$} corresponds to dissimilar outcomes and covariates.}
\label{F:ESS_boxplots_5000}
\end{figure}

 We analyzed each simulated dataset using the proposed transfer learning strategies and summarized the results across replicates. Figure~\ref{F:ESS_boxplots_5000} displays boxplots of ESS for the \textsc{Translate} and Importance Weighting (IW) methods across 250 simulated datasets with a total sample size of $N = 5{,}000$, summarizing scenarios defined by varying degrees of dissimilarity in covariates and outcomes. The corresponding ESS boxplots for $N = 10{,}000$ are provided in Figure~1 in the Supplementary Material.
  We excluded HPP from ESS comparisons because it uses hierarchical shrinkage via mixed-effects models rather than subject-specific weights. Since ESS is defined through the variability of such weights, Stage~1 evaluations focused on \textsc{Translate} and IW, the two methods that explicitly define weighting functions for interpretable comparison.

In the ``dissimilar $Y$ only'' scenario, IW achieved a slightly higher ESS than \textsc{Translate}, though both exceeded the expected anchor size $N\pi_0$. However, a large ESS alone does not imply validity: naive pooling attains the maximum ESS but produces biased anchor estimates. In contrast, \textsc{Translate} maximizes ESS within anchor-aligned pseudopopulations (\ref{pseudo}), preserving unbiasedness and typically achieving higher accuracy than other integrative methods in challenging settings. In the ``dissimilar $X$ and $Y$'' scenario, \textsc{Translate} outperformed IW, whose ESS fell below $N\pi_0$. As predicted by Part~\ref{part2} of Theorem~\ref{Thm1}, \textsc{Translate} remained robust under strong distributional shifts, maintaining ESS well above $N\pi_0$. The \textsc{Translate} ESS closely matched the sum of cohort-level ESS across all settings (Figure~\ref{F:ESS_boxplots_5000}; see also Figure~1 in the Supplementary Material).


Using the subject-specific \textsc{Translate} weights from Stage~1, we applied the weighted estimator in Equation~(\ref{l.hat}) to estimate anchor cohort outcome features, $\mathbb{E}[Y_1 \mid S=0]$, $\mathbb{S}(Y_2 \mid S=0) = \sqrt{\mathbb{V}(Y_2 \mid S=0)}$, and $\mathbb{C}(Y_1, Y_2 \mid S=0)$.
Analytical derivations under the simulation model gave the true values as $0.5$, $\sqrt{371/600}$, and $146/600$, respectively.

\begin{table}[t]
\small
\centering
\begin{tabular}{l|cccc}
\toprule\midrule
\multicolumn{5}{c}{\large \textbf{Absolute Bias and RMSE} ($N=5{,}000$ subjects)} \\
\midrule\midrule

\multicolumn{5}{c}{\textbf{Dissimilar $Y$ Only}} \\
\midrule
\multicolumn{5}{l}{\textit{Absolute Bias}} \\
\midrule
\textbf{Feature} & \textbf{Naive} & \textbf{IW} & \textbf{HPP} & \textbf{\textsc{Translate}} \\
\midrule
$\mathbb{E}[Y_1 \mid S=0]$            & 0.45 (0.00) & 0.25 (0.00) & 0.45 (0.00) & \textbf{0.05} (0.00) \\
$\mathbb{S}[Y_2 \mid S=0]$            & 0.07 (0.00) & 0.07 (0.00) & 0.07 (0.00) & \textbf{0.03} (0.00) \\
$\mathbb{C}(Y_1, Y_2 \mid S=0)$       & 0.29 (0.00) & 0.21 (0.00) & 0.29 (0.00) & \textbf{0.04} (0.00) \\
\midrule
\multicolumn{5}{l}{\textit{RMSE}} \\
\midrule
\textbf{Feature} & \textbf{Naive} & \textbf{IW} & \textbf{HPP} & \textbf{\textsc{Translate}} \\
\midrule
$\mathbb{E}[Y_1 \mid S=0]$            & 0.45 (0.00) & 0.26 (0.00) & 0.45 (0.00) & \textbf{0.08} (0.00) \\
$\mathbb{S}[Y_2 \mid S=0]$            & 0.07 (0.00) & 0.07 (0.00) & 0.07 (0.00) & \textbf{0.05} (0.00) \\
$\mathbb{C}(Y_1, Y_2 \mid S=0)$       & 0.29 (0.00) & 0.21 (0.00) & 0.29 (0.00) & \textbf{0.06} (0.00) \\
\midrule\midrule

\multicolumn{5}{c}{\textbf{Dissimilar $X$ and $Y$}} \\
\midrule
\multicolumn{5}{l}{\textit{Absolute Bias}} \\
\midrule
\textbf{Feature} & \textbf{Naive} & \textbf{IW} & \textbf{HPP} & \textbf{\textsc{Translate}} \\
\midrule
$\mathbb{E}[Y_1 \mid S=0]$            & 1.35 (0.00) & 0.27 (0.01) & 1.35 (0.00) & \textbf{0.04} (0.00) \\
$\mathbb{S}[Y_2 \mid S=0]$            & 0.17 (0.00) & 0.07 (0.00) & 0.17 (0.00) & \textbf{0.03} (0.00) \\
$\mathbb{C}(Y_1, Y_2 \mid S=0)$       & 0.47 (0.00) & 0.18 (0.00) & 0.47 (0.00) & \textbf{0.03} (0.00) \\
\midrule
\multicolumn{5}{l}{\textit{RMSE}} \\
\midrule
\textbf{Feature} & \textbf{Naive} & \textbf{IW} & \textbf{HPP} & \textbf{\textsc{Translate}} \\
\midrule
$\mathbb{E}[Y_1 \mid S=0]$            & 1.35 (0.00) & 0.30 (0.01) & 1.35 (0.00) & \textbf{0.07} (0.00) \\
$\mathbb{S}[Y_2 \mid S=0]$            & 0.17 (0.00) & 0.10 (0.00) & 0.17 (0.00) & \textbf{0.04} (0.00) \\
$\mathbb{C}(Y_1, Y_2 \mid S=0)$       & 0.47 (0.00) & 0.20 (0.00) & 0.47 (0.00) & \textbf{0.05} (0.00) \\
\midrule
\bottomrule
\end{tabular}
\vspace{0.5em}
\caption{Absolute bias and RMSE across two simulation scenarios (means over 250 datasets). Standard errors (500 bootstraps per dataset) are shown in parentheses. Bold indicates the best-performing method(s).}
\label{table:combined-bias-rmse-5000}
\end{table}

We compared \textsc{Translate} with three alternatives, i.e., naive pooled analysis (Naive), inverse weighting (IW), and hierarchical partial pooling (HPP), each applied to all $N$ subjects. Table~\ref{table:combined-bias-rmse-5000} summarizes absolute bias and RMSE for $N = 5{,}000$ subjects. The corresponding results for $N = 10{,}000$ subjects are provided in  Table~S2 in the Supplementary Material. All tables report averages over 250 replications under two scenarios: (i) \textit{Dissimilar $Y$ only} and (ii) \textit{Dissimilar $X$ and $Y$}. 
 Standard errors (in parentheses) were estimated via 500 bootstrap replicates per dataset, and the best-performing methods are highlighted in bold. The results highlight the advantages of the proposed framework. \textsc{Translate} achieved lower bias and RMSE across outcomes and sample sizes, including the difficult ``dissimilar $X$ and $Y$'' case, demonstrating superior stability and precision.

These results demonstrate the robustness of \textsc{Translate} in handling complex covariate-outcome heterogeneity, consistently yielding more accurate and stable inferences than both naive pooling and existing transfer learning methods. Its advantage stems from stabilizing alignment weights rather than optimizing a single predetermined outcome feature.

\section{ Lung Sepsis Outcomes in the Northeast Cohort}\label{S:data analysis}

Our analysis focuses on lung sepsis patients from the Northeastern United States, using data from the eICU Collaborative Research Database described in Section~\ref{S:method}, which includes 6,966 ICU patients nationwide. The Northeast cohort, with 408 patients, is markedly underrepresented, limiting reliable inferences and highlighting potential regional disparities in sepsis management. We examined four clinical markers representing oxygenation (FiO$_2$), kidney function (creatinine), coagulation (platelets), and metabolism (lactate), along with demographic and admission variables to assess sepsis severity. Sex-specific summaries were also evaluated to reveal biological and clinical differences relevant to precision medicine. 

To improve inference for the small Northeast cohort, we applied \textsc{Translate}, leveraging all $N = 6{,}966$ patients while accounting for regional heterogeneity. Stage~1 estimated subject-specific weights via random forests, and Stage~2 produced weighted overall and sex-specific means for the four outcomes.  To evaluate integration performance, we compared five approaches: (i) naive pooling of all $N$ subjects (``Naive''), (ii) analysis of the anchor cohort only (``Northeast Cohort''), (iii) Importance Weighting (``IW''), (iv) Hierarchical Partial Pooling (``HPP''), and (v) \textsc{Translate}. The last three explicitly account for cohort heterogeneity. Inference uncertainty was quantified using $B=500$ bootstrap replicates. 

The \textsc{Translate} population achieved an ESS of 3,140.3 (45.1\%) patients, compared with 1,183.8 (17.0\%) for IW. Across bootstrap samples, ESS differences between \textsc{Translate} and IW were consistently positive, ranging from 600.6 (8.6\%) to 1,886.4 (27.1\%), with a median of 804.8 (11.6\%). These results provide estimand-agnostic evidence of \textsc{Translate}'s superior inferential accuracy. HPP was excluded from ESS comparisons since it uses hierarchical shrinkage rather than subject-specific weighting.

Table~\ref{table:stage2-sepsis-B} summarizes the Stage~2 results for the four clinical outcomes across the five approaches. For each variable, the table reports overall means and sex-specific means for male and female patients in the Northeast. We find that relying solely on the anchor cohort  yields less accurate estimates, particularly for features related to FiO$_2$ levels, which often deviate from patterns revealed by transfer learning methods. While the Naive method tends to produce the smallest standard errors, it likely introduces considerable bias by incorrectly assuming that all four U.S. regions share identical characteristics, an assumption unsupported by prior evidence  \citep{seymour2017epidemiology}. 
This bias is evident in the markedly different estimates for certain outcomes produced by the Naive method compared to other approaches. The most precise transfer learning approaches, identified by the lowest standard errors, are highlighted in bold. Among all approaches, \textsc{Translate} consistently delivers the most accurate and precise inferences across the univariate outcomes. 
Figure~2 in the Supplementary Material provides a graphical summary showing the estimated means and 95\% confidence intervals.

\begin{table}[htbp]
\small
\centering
\renewcommand{\arraystretch}{1.5} 
\resizebox{\textwidth}{!}{
\begin{tabular}{|l|c|c|c|c|c|}
\hline
\rowcolor{lightgray!15}\multicolumn{6}{|c|}{\rule{0pt}{3ex}\large \textbf{Lactate}} \\
\hline
\textbf{Feature} & \textbf{Naive} & \textbf{Northeast Cohort} & \textbf{IW} & \textbf{HPP} & \textbf{TRANSLATE} \\
\hline
Overall Mean  & 3.28 (0.04) & 3.47 (0.16) & 3.30 (0.14) & 3.46 (0.13) & 3.37 (\textbf{0.08}) \\ 
Mean for Males  & 3.30 (0.05) & 3.37 (0.17) & 3.23 (0.16) & 3.47 (0.13) & 3.41 (\textbf{0.09}) \\
Mean for Females  & 3.25 (0.06) & 3.63 (0.33) & 3.42 (0.18) & 3.44 (0.15) & 3.32 (\textbf{0.11}) \\ 
\hline
\rowcolor{lightgray!15}\multicolumn{6}{|c|}{\rule{0pt}{3ex}\large \textbf{Platelets}} \\
\hline
\textbf{Feature} & \textbf{Naive} & \textbf{Northeast Cohort} & \textbf{IW} & \textbf{HPP} & \textbf{TRANSLATE} \\
\hline
Overall Mean  & 168.90 (1.24) & 158.49 (4.39) & 163.56 (3.51) & 162.10 (3.65) & 165.82 (\textbf{2.18}) \\
Mean for Males   & \underline{162.26} (1.75) & 156.61 (7.50) & 159.94 (5.23) & \underline{156.40} (3.82) & \underline{160.56} (\textbf{2.83}) \\ 
Mean for Females  & \underline{176.75} (1.48) & 161.27 (6.27) & 168.94 (4.30) & \underline{170.68} (3.90) & \underline{172.99} (\textbf{3.06}) \\ 
\hline
\rowcolor{lightgray!15}\multicolumn{6}{|c|}{\rule{0pt}{3ex}\large \textbf{Creatinine}} \\
\hline
\textbf{Feature} & \textbf{Naive} & \textbf{Northeast Cohort} & \textbf{IW} & \textbf{HPP} & \textbf{TRANSLATE} \\
\hline
Overall Mean & 2.00 (0.02) & 2.04 (0.08) & 2.04 (0.11) & 2.03 (\textbf{0.04}) & 2.02 (\textbf{0.04}) \\ 
Mean for Males  & \underline{2.18} (0.03) & \underline{2.21} (0.11) & 2.16 (0.14) & \underline{2.18} (\textbf{0.04}) & \underline{2.17} (\textbf{0.04}) \\ 
Mean for Females   & \underline{1.80} (0.02) & \underline{1.79} (0.11) & 1.86 (0.09) & \underline{1.80} (\textbf{0.04}) & \underline{1.81} (\textbf{0.04}) \\ 
\hline
\rowcolor{lightgray!15}\multicolumn{6}{|c|}{\rule{0pt}{3ex}\large \textbf{FiO$_2$}} \\
\hline
\textbf{Feature} & \textbf{Naive} & \textbf{Northeast Cohort} & \textbf{IW} & \textbf{HPP} & \textbf{TRANSLATE} \\
\hline
Overall Mean  & 56.30 (2.58) & 63.65 (12.93) & 53.82 (6.42) & 59.86 (7.87) & 48.35 (\textbf{2.61}) \\ 
Mean for Males  & 57.94 (3.61) & 58.50 (17.43) & 53.13 (8.66) & 61.37 (8.04) & 48.32 (\textbf{4.57}) \\
Mean for Females  & 54.36 (4.45) & 71.32 (32.36) & 54.98 (9.85) & 57.53 (9.33) & 48.38 (\textbf{3.64}) \\ 
\hline
\end{tabular}
}
\vspace{0.5em}
\caption{
{The estimated means and corresponding standard errors (in parentheses) for three target cohort features, organized by different clinical outcomes (blocks of rows) and five competing strategies (columns). Within each clinical outcome feature, the transfer learning  approach with the smallest standard error is emphasized in bold. For each clinical outcome and analytical approach, sex-specific means that show a significant difference between males and females are underlined. ``Overall Mean'', ``Mean for Males'', and ``Mean for Females'' represent the target cohort features $\mathbb{E}[Y \mid S=0]$, $\mathbb{E}[Y \mid S=0, \,\text{male}]$, and $\mathbb{E}[Y \mid S=0, \,\text{female}]$, respectively. See Section \ref{S:data analysis} for a detailed discussion. 
}
}
\label{table:stage2-sepsis-B}
\end{table}

While Inverse Weighting (IW) can, in principle, adjust for cohort heterogeneity, it often yields suboptimal weighting in practice, as reflected by the consistently more precise estimates from \textsc{Translate}.  The difference arises from their distinct weighting strategies. Figure~3 of the Supplementary Material shows that IW places roughly half of the total weight on the anchor cohort, whereas \textsc{Translate} reallocates 91.7\% to external cohorts, distributing weights based on their similarity in both covariates and outcomes. This targeted reweighting enables more effective and reliable information transfer from external data.

Both HPP and \textsc{Translate} identify significant sex-based differences in creatinine levels within the Northeast cohort, consistent with established evidence that higher baseline creatinine in males reflects greater muscle mass and influences renal function assessment in sepsis patients \citep{levey1999gfr}. While platelet count is an established marker of sepsis severity, sex-related differences are less documented \citep{koyama2018time} and not significant in the small Northeast cohort alone; however, by integrating external data, HPP and \textsc{Translate} enhance precision and reveal significant male-female differences in platelet counts.

\begin{figure}
    \centering
    \includegraphics[width=\textwidth]{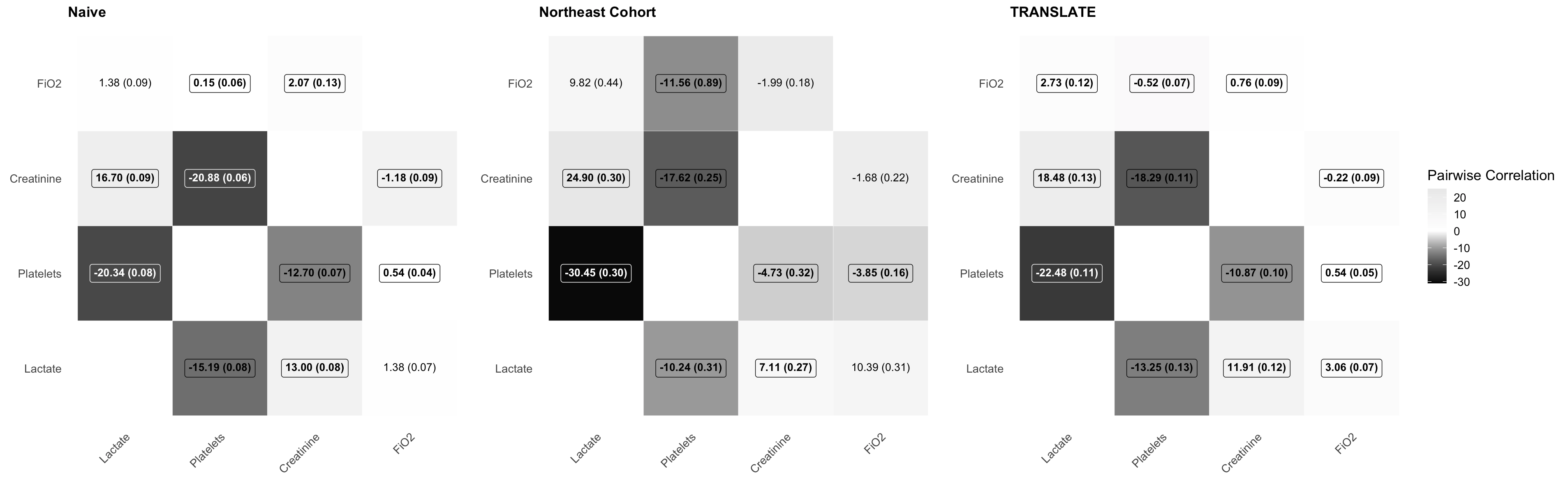}
    \caption{Estimated pairwise correlations (and standard errors) multiplied by 100 (percentage points), by method and sex. Within each method, the upper triangle represents values of females, while the lower triangle represents values of males. Sex-specific estimates that differ significantly (male vs.\ female) are bolded and framed.}
    \label{fig:pairwise_cor}
\end{figure}

Figure~\ref{fig:pairwise_cor} shows the estimated pairwise correlations among lactate, platelets, creatinine, and FiO$_2$ for male and female sepsis patients in the Northeast. IW and HPP were excluded since they are not designed for  correlation analyses of multiple outcomes. Across all biomarker pairs, \textsc{Translate} achieved the most precise or comparable estimates, as quantified by standard errors.
 Its estimates consistently fell between those from the biased but precise Naive pooled analysis and the unbiased but imprecise Northeast Cohort analysis. \textsc{Translate} also detected significant sex-specific correlation differences, underscoring its sensitivity and efficiency in capturing clinically relevant heterogeneity within small regional cohorts.

To evaluate the correlations, we performed pairwise bootstrap comparisons between \textsc{Translate} and alternative methods (Table~S2 in the Supplementary Material).
 Compared with the Naive approach, \textsc{Translate} differed significantly for four of six biomarker pairs: (lactate, creatinine), (lactate, FiO$_2$), (platelets, creatinine), and (platelets, FiO$_2$), and for three of six pairs relative to the Northeast Cohort analysis. These results suggest that \textsc{Translate} captures distinct and more reliable correlation structures, particularly for relationships involving creatinine and FiO$_2$.

 These results show that \textsc{Translate} effectively integrates external data to improve efficiency and maintain valid inferences for Northeast patients. By addressing regional variability and sex-based differences, it enhances sepsis analysis for underrepresented groups and supports tailored management strategies \citep{pollard2018eicu}.

\section{Discussion}\label{S:discussion}

Drawing reliable inferences for small anchor cohorts is challenging due to limited sample sizes. Transfer learning can improve precision by leveraging larger external datasets, but integration is complicated by heterogeneity in covariates and outcomes. We proposed \textsc{Translate}, an adaptive weighting framework that constructs an anchor-aligned pseudopopulation maximizing composite effective sample size while aligning covariate and outcome distributions across cohorts. This principled alignment enables efficient, unbiased information transfer and flexible estimation of diverse multivariate functionals (means, variances, covariances, correlations) and subgroup comparisons. Theoretically, \textsc{Translate} attains higher precision than anchor-only analyses while down-weighting dissimilar cohorts, ensuring unbiasedness and improved precision even under substantial heterogeneity.  
In the lung sepsis application, \textsc{Translate}  demonstrates stable performance despite substantial regional heterogeneity and a small anchor cohort. By reallocating weight toward external cohorts most similar to the Northeast population, the method achieves improved precision for means and correlations while avoiding the bias of naive pooling.

A key feature of \textsc{Translate}  is its estimand-agnostic design. Rather than tailoring weights to a single outcome or regression parameter, the method constructs a pseudopopulation aligned with the anchor cohort in joint covariate–outcome space, enabling valid estimation of a broad class of multivariate functionals, including means, variances, correlations, and subgroup contrasts. This flexibility distinguishes \textsc{Translate}  from outcome-specific transfer or calibration approaches and is particularly valuable in exploratory or multifeature settings, such as biomarker profiling in underrepresented populations. In the meantime, the  framework relies on estimating cohort membership probabilities, equivalently density ratios, to construct alignment weights. When these probabilities are modeled parametrically, for example via multinomial logistic regression, correct model specification is required for exact identification of the target pseudopopulation. We emphasize that this assumption is shared by much of the existing importance-weighting and transportability literature and should be interpreted as a working model rather than a guarantee. In practice, misspecification of the cohort probability model can induce bias and inflate variance, particularly when anchor and external cohorts differ markedly.

To mitigate this risk, our framework allows the use of flexible machine-learning methods, such as random forests, boosting, or Bayesian additive regression trees, for estimating cohort probabilities. These methods relax parametric assumptions and can better capture complex, high-dimensional differences across cohorts. However, their use introduces additional theoretical challenges. In particular, nonparametric or ensemble learners typically converge at rates slower than root-n, and such rates may propagate into the second-stage weighted estimators. Our asymptotic results implicitly require sufficiently fast convergence of the estimated alignment weights so that first-stage uncertainty does not dominate the limiting distribution. While this condition is standard in the weighting literature, it may be restrictive in some high-dimensional regimes.

A possible remedy is to augment the current weighting strategy with outcome modeling, yielding doubly robust or augmented estimating equations. A substantial body of prior research shows that combining inverse weighting with outcome regression can simultaneously improve efficiency and reduce sensitivity to misspecification of the weighting model. In our context, augmentation could stabilize inference when density-ratio estimation is noisy and potentially restore root-n convergence under weaker conditions on first-stage learners. Developing such augmented versions would also strengthen connections to semiparametric efficiency theory and modern causal transportability methods.

Therefore, several extensions merit further investigation. First, formalizing augmented or doubly robust versions of \textsc{Translate}   would strengthen both robustness and efficiency. Second, sharper theoretical results that explicitly characterize the impact of first-stage convergence rates on second-stage inference would clarify when flexible learners can be safely deployed. Third, extensions to federated or privacy-restricted settings, where only summary information is available from external cohorts, would broaden applicability. Finally, high-dimensional outcome settings, such as genomics or imaging, raise additional challenges for weight stability that may benefit from regularization or representation learning.


\section*{Acknowledgments}
This work was supported by the National Institutes of Health under awards CA249096  (YL),  and  CA269398 and CA209414  (SG and YL). 

\section*{Data Availability Statement}
The data used in this paper are available in the  \href{https://eicu-crd.mit.edu/}{eICU Collaborative Research Database}. Researchers seeking to use the database must request access; details at \href{https://eicu-crd.mit.edu/gettingstarted/access/}{https://eicu-crd.mit.edu/gettingstarted/access/}. 

\section*{Supplementary material}
Supplementary material for ``Enhancing Inference for Small Cohorts via Transfer Learning and Weighted Integration of Multiple Datasets". This material contains summaries of the clinical, demographic, and biomarker variables from the eICU dataset, the theorem proofs, and  additional figures and tables, including extended simulation results and secondary analyses from the lung sepsis application.


\bibliographystyle{apalike} 
\bibliography{AllRefs_mar2022}


\end{document}